\documentclass[prl,aps,twocolumn,showpacs,superscriptaddress,floatfix]{revtex4}
\usepackage{graphicx}
\usepackage{bm}

\usepackage[dvips]{color}
\usepackage{graphicx}

\begin{document}

\title{Hubbard Model on the Pyrochlore Lattice: a 3D Quantum Spin 
Liquid}

\author{B. Normand}
\affiliation{Department of Physics, Renmin University of China, Beijing
100872, China}

\author{Z. Nussinov}
\affiliation{Department of Physics, Washington University, St. Louis,
MO 63160, U.S.A.}

\date{\today}

\begin{abstract}
We demonstrate that the insulating one-band Hubbard model on the pyrochlore 
lattice contains, for realistic parameters, an extended quantum spin-liquid 
phase. This is a three-dimensional spin liquid formed from a highly degenerate 
manifold of dimer-based states, which is a subset of the classical dimer 
coverings obeying the ice rules. It possesses spinon excitations, which are 
both massive and deconfined, and on doping it exhibits spin-charge separation. 
We discuss the realization of this state in effective $S = 1/2$ pyrochlore 
materials.
\end{abstract}


\maketitle

The quantum spin liquid \cite{rafa} has become the focal point for our 
understanding of many of the most fundamental issues in strongly correlated 
systems. These include exotic quantum phases, quantum critical physics, the 
relevance of broken symmetries, topological order, entanglement, and the 
possibly fractional nature of elementary excitations in both gapped and gapless 
states \cite{rb}. The search for theoretical realizations of these ideas has 
led to numerous proposed models, which while highly informative have generally 
been too simple or abstract to apply to real materials \cite{qslmodels}. The 
search for materials realizations is a very active field where much current 
attention is focused on kagome systems \cite{qslkagome}, triangular organics 
\cite{qslto}, and other frustrated $S = 1/2$ and $S = 1$ quantum magnets. 
However, materials complexities such as impurities, Dzyaloshinskii-Moriya 
interactions, spin-orbit coupling, and other anisotropies in real and spin 
space have to date caused strong departures from theoretical ideals. 

Frustrated quantum magnets offer one of the most promising routes to 
spin-liquid behavior \cite{rafa,rb}. Frustration presents a formidable 
barricade to theoretical understanding, because the ground manifold is 
quite generally a set of highly degenerate basis states, with little or 
no separation emerging in an exact treatment of the interactions \cite{rn}. 
Numerical calculations converge very slowly due to this proliferation of 
near-ground states \cite{hard}. Fluctuations in such a manifold may lead 
to a range of exotic phenomena \cite{Fradkin-book,TQO,DQCP}, and the 
departures mentioned above are strong because any perturbation is strongly 
relevant in a highly degenerate system. Few exact results are available, 
although these afford essential insight \cite{baxter,Klein82,Chayes89,rs}. 

In this Letter, we discuss the one-band Hubbard model, showing that on a 
half-filled pyrochlore lattice it gives a highly frustrated intratetrahedral 
spin model with only weak perturbations. This model contains an exactly 
solvable Klein point, about which there is an extended region of parameter 
space where the ground state is a three-dimensional (3D) quantum spin 
liquid. This state hosts massive spinon excitations, which are deconfined 
and move in all three dimensions within the lattice. The parameter range for 
the spin-liquid phase lies exactly in the regime of many magnetic materials.

\begin{figure}[t]
\includegraphics[width=8.5cm]{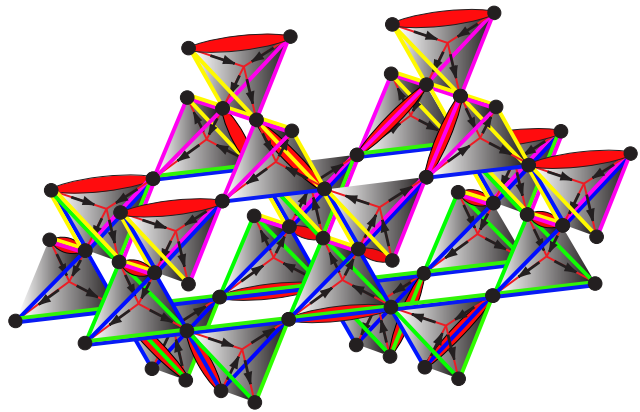}
\caption{(color online) Pyrochlore lattice. The magnetic ions (black circles) 
form a 3D array of corner-sharing tetrahedra.}
\label{fig1}
\end{figure}

The pyrochlore lattice, shown in Fig.~1, is a 3D array of corner-sharing 
tetrahedra, has cubic symmetry, and is a geometry widespread in 
transition-metal and rare-earth oxides. Most such materials have 
half-filled bands and are Mott-Hubbard insulators due to their interactions. 
We begin with the Hubbard model,
\begin{equation}
H_{\rm Hubb} = - t \, \mbox{$\sum_{\langle ij \rangle, \sigma}$}  c_{i\sigma}^{\dag}
c_{j\sigma} + U \, \mbox{$\sum_{i}$} n_{i\uparrow} n_{i\downarrow},
\label{eeh}
\end{equation}
where $c_{i \sigma}^{\dag}$ creates an electron at site $i$ with spin $\sigma$ 
and $n_{i\sigma} = c_{i\sigma}^{\dag} c_{i\sigma}$ is the number operator. We use 
it to discuss the spin liquid with no theoretical abstractions and with 
controlled approximations. A perturbative expansion in $t/U$ for the 
half-filled band leads to 
\begin{equation}
H = H_t + J_3 \, \mbox{$\sum_{\langle \langle i j \rangle \rangle}$} \vec{S}_{i} 
\cdot \vec{S}_{j} +  {\cal{O}} ( t^{6}/ U^{5} ),
\label{eehe}
\end{equation}
where $\langle \langle i j \rangle \rangle$ denotes next-neighbor site pairs 
and
\begin{equation}
H_t = \mbox{$\sum_l$} \, [ {\textstyle \frac12} J_1 {\bf S}_{l, \rm tot}^2 
+ {\textstyle \frac{1}{4}} J_2 {\bf S}_{l, \rm tot}^4 ]
\label{eht}
\end{equation}
is a sum of purely intratetrahedral spin interactions written in terms of 
the total spin ${\bf S}_{l, \rm tot} = {\bf S}_{l1} + {\bf S}_{l2} +  {\bf S}_{l3} 
+  {\bf S}_{l4}$ on each tetrahedron, $l$ \cite{rnbnt}. In Eqs.~(\ref{eehe}) 
and (\ref{eht}), 
\begin{eqnarray}
J_{1} & = & 4t^{2}/U - 160 t^{4}/U^{3} + {\cal{O}} (t^{6}/U^{5}), \label{ej} 
\\ J_{2} & = & 40 t^{4}/U^{3} + {\cal{O}} (t^{6}/U^{5}), \;\;\; J_{3} \; = \; 
4t^{4}/U^{3} + {\cal{O}} (t^{6}/U^{5}), \nonumber
\end{eqnarray}
the very large prefactors in $J_1$ and $J_2$ arising from the many permutations 
of fourth-order processes within the tetrahedron. Thus $J_{3} \ll J_{1,2}$ 
and to an excellent approximation one has an intratetrahedral Hamiltonian
$H_t$, with only weak interactions coupling spins in different tetrahedra. 

$H_t$ (\ref{eht}) has a unique point for one particular parameter ratio, 
$J_2 = J_{2c} = - J_1$, occurring when $t/U = 1/\sqrt{30}$, where all singlet 
and triplet states of the four spins on each tetrahedron have energy zero 
\cite{sm}. Thus any tetrahedon containing a dimer, a singlet ($S = 0$) state 
of any two spins, and represented by red ellipses in Figs.~1 and 2, has energy 
zero. Further, because the number of dimers equals the number of tetrahedra on 
the pyrochlore, all states of the whole system with precisely one dimer per 
tetrahedron (Fig.~1) are exact, zero-energy ground states. The set of classical 
dimer coverings maps exactly to the six-vertex model, represented by the six
possible ``two-in, two-out'' configurations of the black arrows in Figs.~2(a) 
and (b), and hence to the ice problem. Pauling deduced an exponential lower 
bound on the number of states in this ground manifold, $N_g > (3/2)^{N/2}$ with 
$N$ the system volume (number of tetrahedra) \cite{rp}, proving that it has 
extensive degeneracy. This is a Klein point \cite{Klein82,sm}. It is a dimer 
liquid with algebraic correlations, a classical critical point at which all 
(of the exponentially many) states are connected by local dimer fluctuations. 
The excitations created by breaking a dimer are two massive spinons, which 
propagate freely \cite{rnbnt}. 

\begin{figure}[t]
\includegraphics[width=7.5cm]{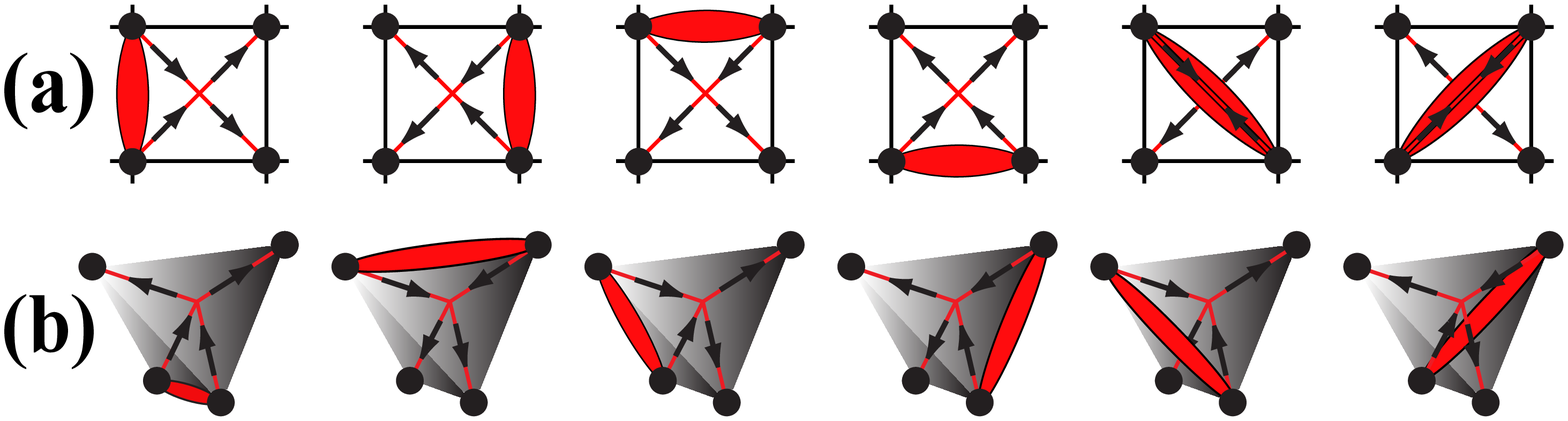} \\
\includegraphics[width=3.5cm]{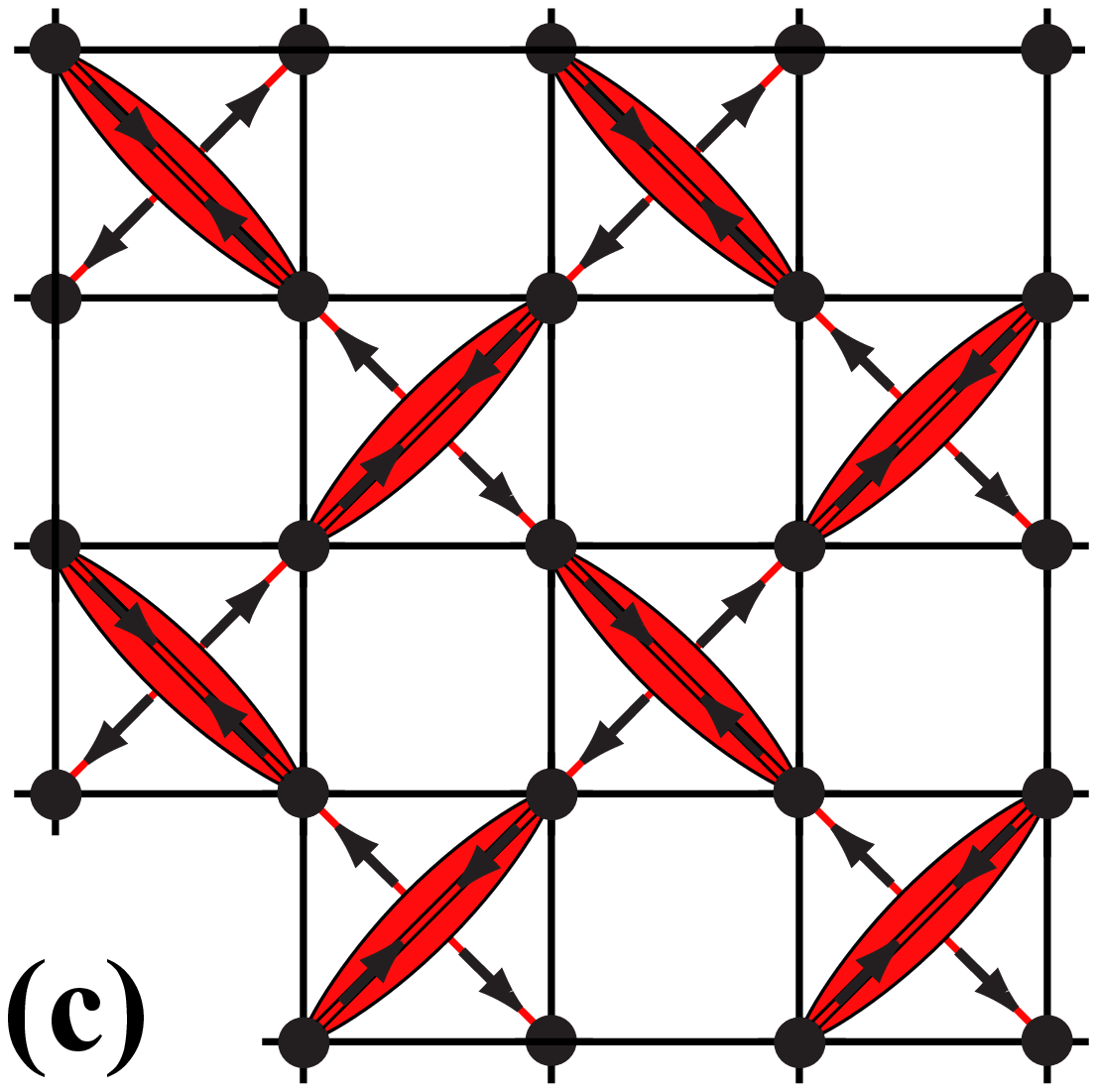}\hspace{0.1cm}\includegraphics[width=3.9cm]{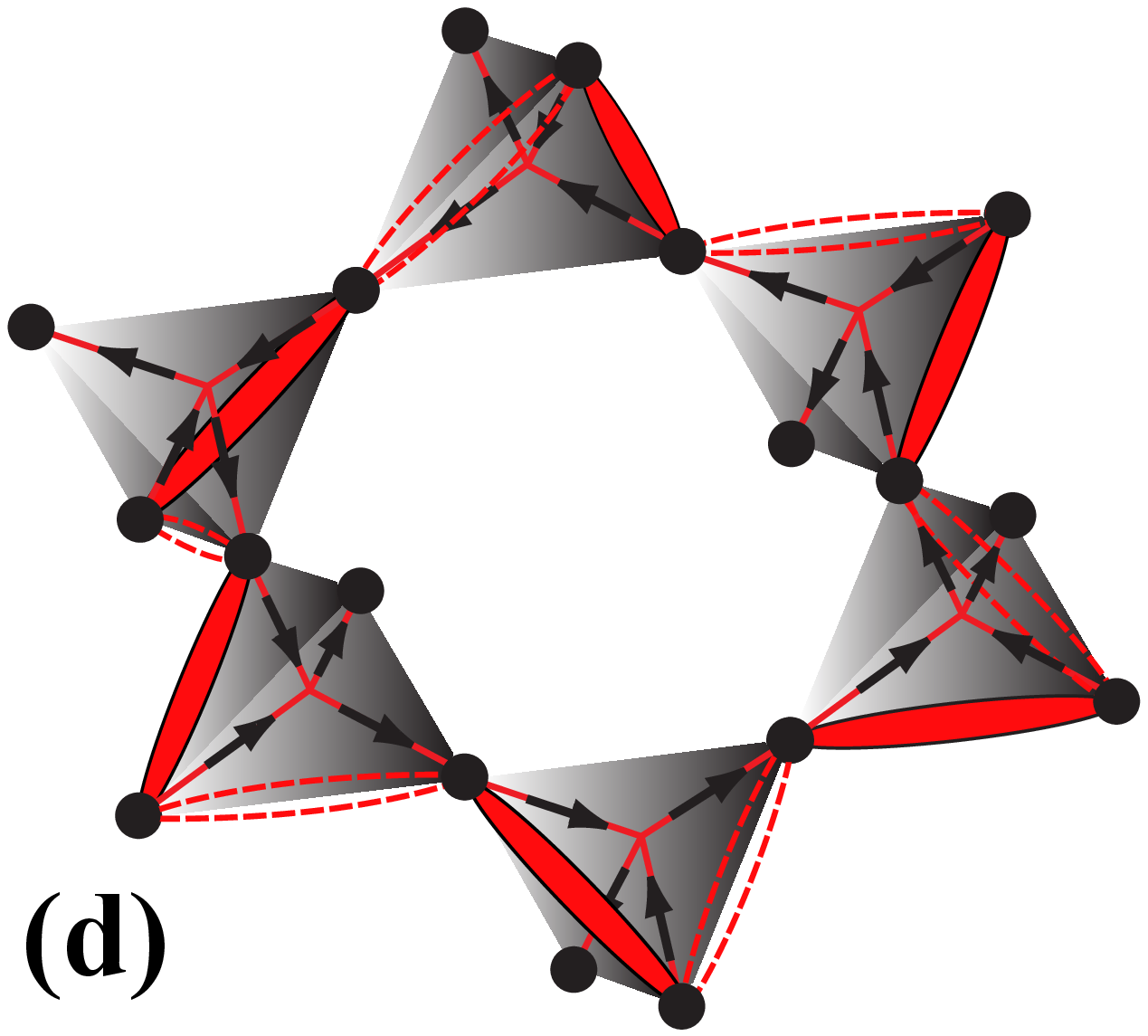} 
\caption{(color online) One-dimer configurations on a single tetrahedron and 
their six-vertex representations (a) in 2D and (b) in 3D. (c) The maximally 
flippable state (all-diagonal dimer covering) of the checkerboard lattice, 
which requires only two vertex types. (d) One hexagon of the pyrochlore 
lattice with the dimers of the surrounding tetrahedra in a flippable 
configuration, as shown by the black arrows; dashed red ellipses indicate 
the flipped dimer state.}
\label{fig2}
\end{figure}

However, it is unrealistic to expect any physical system to be exactly at 
the Klein point. To understand which of its many properties may be preserved 
in a real material, it is essential to analyze the effect of perturbations. 
Because every site must be part of one dimer, the quantum mechanical 
fluctuations of the dimer liquid are local rearrangement processes on a 
closed path. Figure 2(d) illustrates the minimal possible dimer rearrangment 
on the pyrochlore lattice and Fig.~3 shows more generally how all such 
processes may be described by loops, which represent the overlap of the 
two dimer coverings connected by the fluctuation. Here we extend the 
loop-graph analysis of Ref.~\cite{rnbnt} to 3D to deduce the nature of the 
pyrochlore ground state close to the Klein point. A perturbation $\Delta H
 = \sum_{ij} {\Delta J} \, {\bf S}_i {\bf \cdot S}_j$ allows us to analyze 
exactly all the leading physically relevant terms in the pyrochlore 
Hamiltonian. Deviations from the Klein-point ratio, resulting from 
alterations to $J_1$ or $J_2$ in Eq.~(\ref{eht}), are represented exactly 
by considering nearest-neighbor sites $\langle ij \rangle$, and deviations 
from $H = H_t$, particularly the $J_3$ terms in Eq.~(\ref{ej}), by using 
next-neighbor sites $\langle\langle ij \rangle\rangle$. 

Loops, or dimer fluctuations, exist on all length scales, but as we show 
below the most important contributions are made by short loops, which 
describe local processes. The very shortest loops in the 3D (2D) pyrochlore 
lattice are 12- (8-)bond paths around a single hexagon [Fig.~2(d)] (vacant 
square [Fig.~2(c)]). Rearranging the 6 (4) dimers corresponds to flipping 
the sign of the arrow in the six-vertex representation, and we refer to 
local dimer configurations allowing these loops as ``flippable plaquettes'' 
[Figs.~2(c) and (d)]. These ``Rokhsar-Kivelson'' (RK) loops are zero-energy 
processes \cite{rnbnt}. Nevertheless, dimer coverings with maximal numbers 
of flippable plaquettes also maximize longer contributing loops and thus 
form the basis for the new ground states in the presence of a perturbation. 
In 2D, two of the vertices are special in that all four edges of the square 
have an ``in-out'' arrow configuration [Fig.~2(a)], such that regular arrays 
of these two can make every plaquette flippable [Fig.~2(c)]. The degeneracy 
of the submanifold of ``maximally flippable'' states is then ${\cal O}(1)$, 
and the ground states on both sides of the Klein point are valence-bond 
crystals, with a preferred static dimer order \cite{rnbnt}. 

\begin{figure}[t]
\includegraphics[width=8.0cm]{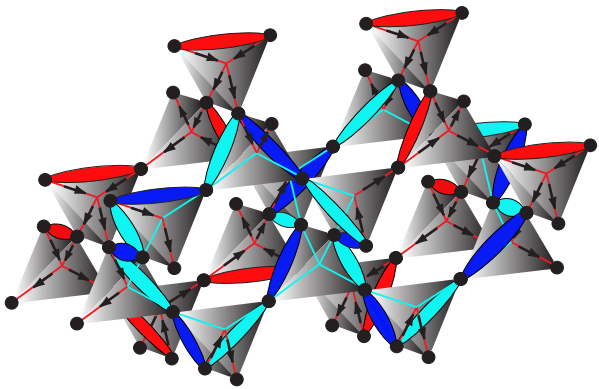} 
\caption{(color online) Dimer fluctuation process, shown as a loop of 
alternating light and dark blue dimers. This is a 22-bond loop (see text).}
\label{fig3}
\end{figure}

This result is a special property of the six-vertex model in a square geometry, 
and the situation in 3D is dramatically different. All six vertices are 
equivalent [Fig.~2(b)] and every tetrahedron has two edges destroying the 
flippability of the four associated hexagons \cite{sm}, making clear that not 
all hexagons in the 3D pyrochlore can be flippable. Our proof of spin-liquid 
nature around the Klein point is the demonstration i) that the ground manifold 
has massive degeneracy and ii) that this degeneracy is unbroken by any of the 
leading quantum fluctations. 

Flippable hexagons can be counted by considering the four interlocking kagome 
(111) planes of the pyrochlore lattice, highlighted in different colors in 
Fig.~1. The maximally flippable dimer coverings have bilayers of tetrahedra 
ensuring maximal flippability of the hexagons in two of the kagome planes, 
shown in blue and green in Fig.~1, interleaved with equivalent bilayers 
maximizing the other two planes (yellow and purple). The maximal number 
density of flippable hexagons is 1/3 \cite{sm}. Tetrahedra in the layer 
between the bilayers retain a two-fold degree of freedom in singlet 
orientation (Fig.~1), equivalent to the relative arrow direction between 
bilayers. The degeneracy of the maximally flippable manifold is then 
$N_f = 9$$\times$$2^{L/3}$ \cite{sm}, where the system volume is $N = L^3/4$ 
and $L$ is the linear dimension. 

\begin{table}[b]
\caption{Lowest-order loops in the pyrochlore lattice \cite{sm}.}
\begin{center}
\begin{tabular*}{8.5cm}{@{\extracolsep{\fill}} c c c c c c c c c}
\hline \hline
Loop Length  & 12$\,$ & 16$\,$ & 16$\,$ & 20$\,$ & 20$\,$ & 22$\,$ & 24$\,$ 
& 26$\,$ \\
$\Delta H_{ab}$  & 0$\,$ & 0$\,$ & $\frac{\Delta J}{128} \,$ & 0 & $- 
\frac{\Delta J}{512} \,$ & $- \frac{\Delta J}{1024} \,$ & $- \frac{\Delta 
J}{2048} \,$ & $- \frac{\Delta J}{4096} \,$ \\ 
Loop Density & 1 & $\frac{1}{2} \,$ & $\frac{1}{2} \,$ & 1 & 1 & 8 $\,$ & 
$\frac{1}{2} \,$ & 2 $\,$ \\ 
\hline \hline
\end{tabular*}
\end{center}
\end{table}

To determine the ground manifold in the presence of physical perturbations, 
we evaluate the matrix elements of $\Delta H$ for each loop type; specifically, 
we compute $\Delta H_{ab} = \langle \psi_a | \Delta H | \psi_b \rangle$ for 
states $|\psi_a \rangle$ and $|\psi_b \rangle$ differing by one loop of dimers 
(Fig.~3). The importance of the maximally flippable configurations, anticipated 
above, is proven by considering all loops on the pyrochlore lattice involving 
2 or 3 hexagons and generated by a single RK defect. The calculations are 
presented in the Supplemental Material \cite{sm} and the results summarized 
in Table I. Beyond the RK loop, the size of the matrix elements clearly falls 
exponentially with loop length, demonstrating the key role of the shortest 
loops. The ground manifold is therefore composed of all maximally flippable 
states containing precisely one RK defect, which may be on any one of the $N/3$ 
flippable hexagons, and hence the dimension of this manifold of basis states, 
$N_p = 3N$$\times$$2^{L/3}$, is massive and exponential in $L$ \cite{sm}. 

To construct the ground-state wave function from this manifold we require 
the loop density, or number of each loop type per flippable hexagon. While 
one type of 16-bond loop process contributes the most energy, the highest 
densities are found (Table I) for 22-bond loops, which correspond to flipping 
dimers around two hexagons sharing opposite edges of a single tetrahedron 
(Fig.~3). Unlike the 2D case, in 3D the lowest-order loops do interfere, and 
the deciding quantum fluctuations are the three shortest contributing loops 
in Table I \cite{sm}. By considering the most general linear combinations, 
$|\psi \rangle = \sum_a c_a |\psi_a \rangle$, of states $|\psi_a \rangle$ 
based on the maximally flippable configurations with a single RK defect, we 
find that there are $N_f$ distinct ground-state wave functions optimizing 
the loop (dimer fluctuation) contribution and that each of these is a 
superposition of ${\cal O}(N)$ basis wave-function pairs with equal amplitudes 
$|c_a|$ \cite{sm}. The sign of $\Delta J$ causes differences not only in 
the phase structure of the variational wave functions (the signs of the 
coefficients $\{c_a\}$) \cite{rnbnt,sm} but also in their energies; we obtain 
$\Delta E = - {\textstyle \frac{1}{256}} u^2 N \Delta J$ for $\Delta J > 0$ and 
$\Delta E = {\textstyle \frac{1}{128}} v^2 N \Delta J$ for $\Delta J < 0$, 
where $u,v \simeq 1$ are normalization coefficients. Because 22-bond loops 
connect hexagons in neighboring bilayers, they can be used to illustrate a 
final, crucial property. There is no preferred direction for circumscribing 
a hexagon and all such loops contribute the same energy for either maximally 
flippable dimer covering of each bilayer. By extension to any type of 
interbilayer loop \cite{sm}, there is no mechanism to lift the bilayer 
degeneracy, and hence all $N_p$ states in the ground manifold form the same 
type of minimum-energy state. 

To summarize, our loop calculations verify that a highly degenerate ground
manifold persists under physical perturbations away from the Klein point. 
Further, all states in this manifold gain energy from mutual resonance. 
Linear combinations of these states span all dimensions and break no lattice 
symmetries. These are the qualitative energetic and spatial criteria for a 
spin liquid. However, a proof of quantum spin-liquid nature as a strict 
zero-temperature statement requires specific topological criteria. In the 
ground manifold of the non-Klein-point model, local loop processes reflect 
local gauge-type symmetries and their matrix elements determine the $N_f$ 
variational ground states. These states are linked by $O(L^2)$ local processes, 
which correspond to system-scale, planar (dimension $d = 2$) loops, reflecting 
non-local ``emergent'' Z$_2$ symmetries associated with each bilayer \cite{sm}. 
In systems of finite size, this symmetry is broken but the accompanying 
spectral gaps are exponentially small in $L^2$. The associated topological 
degeneracy in this and other models with $d \ge 1$ processes \cite{rno} 
is analogous to 2D quantum dimer models where the ground states are also 
equal-amplitude superpositions of dimer coverings. These models have non-local 
$d = 1$ symmetries associated with the parity of the even or odd number of 
dimers cut by 1D loops around the entire system, and are understood as 
gapped quantum liquids with Z$_2$ topological order \cite{Fradkin-book}. 
The extension of these topological concepts to dimer states in 2D systems 
of real $S = 1/2$ quantum spins has been demonstrated in recent detailed 
calculations \cite{rpspgc,rws}. Our analysis provides several key 
additional ingredients to this discussion \cite{sm}, proving rigorously 
from energetic and topological criteria that the pyrochlore spin model 
possesses zero-temperature quantum spin-liquid behavior in 3D over an 
extended region of the non-Klein-point parameter space, as represented 
in Fig.~4. 

\begin{figure}[t]
\includegraphics[width=7.0cm]{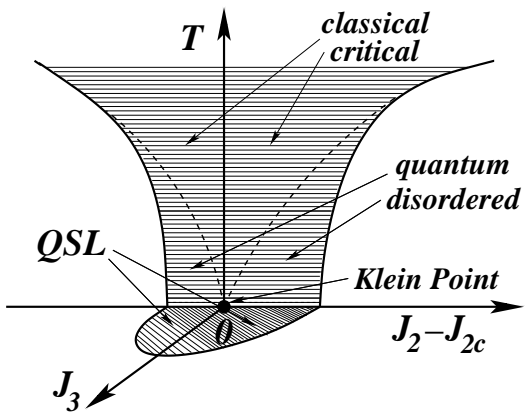}
\caption{(color online) Schematic phase diagram for the quantum spin liquid 
(QSL) phase of the $S = 1/2$ pyrochlore. $J_2 - J_{2c}$ and $J_3$ represent 
respectively the intra- and intertetrahedron perturbations away from the Klein 
point. Solid lines indicate phase transitions and dashed lines crossovers.} 
\label{fig4}
\end{figure}

The basis states of this spin liquid are a subset of the Klein-point 
dimer coverings, and hence are ``divergence-free'' in the six-vertex arrow 
representation (Fig.~2). It has been argued by analogy with continuum 
Gaussian electrostatics that spin-ice systems can be described by a U(1) 
gauge field theory \cite{sm}. However, for our microscopic model, where all 
states in the ground manifold are known exactly, as are all loop processes 
connecting them (Table I), we find that the local dimer rearrangements are 
more complex than those of a U(1) gauge theory alone. Independent of an 
approximate field-theory description, the variational ground states we have 
constructed at finite $\Delta J$ are an exact quantum spin liquid, meaning 
that states in the ground manifold are no longer individual eigenstates of 
the non-Klein-point Hamiltonian and are connected by quantum fluctuations 
to all other states in the same topological sector. 

In the schematic phase diagram of Fig.~4, the solid lines indicate phase 
transitions, which may be of first or second order; different types of 
ground state become increasingly competitive as the intratetrahedron 
singlet-triplet energy splittings increase, but in contrast to 2D a 
transition requires a finite separation from the Klein point. The dashed 
lines indicate the energy scales for a thermally driven crossover from 
quantum to classical behavior, occurring when the temperature exceeds the 
splitting of the Klein-point manifold and the physics of the ``Coulomb phase'' 
of spinons is restored \cite{rnbnt}. The absolute value of this splitting is 
is remarkable. From the matrix elements in Table I it is two orders smaller 
even than the perturbation $\Delta J$, so that, even far from the Klein 
point, the entire manifold would be split on an energy scale well below 1 K 
for any real material. The practical criterion for spin-liquid nature is that 
no local probe can discern any type of order. At finite temperatures, any 
expectation value is a Gibbs average, a sum over exponentially many states 
with small (or vanishing) energy splittings, and thus will vanish for 
temperatures above (or below) the crossover \cite{sm}. 

Armed with a microscopic model for a quantum spin-liquid state, we consider 
the nature of its excitations. High-dimensional fractionalization of both 
spin \cite{rnbnt} and charge \cite{charge} degrees of freedom has been 
considered before in pyrochlore-based geometries. A spin excitation is 
the destruction of a dimer to create one defect tetrahedron (DT) and two 
free spins [Fig.~5(a)]. The finite energy cost for this process means 
these are massive spinons, with $m_s = 15 J_2/16$ \cite{sm}. From the 
dimensional reduction \cite{sm} in our quantum spin liquid, the spinons 
are constrained to move on lines \cite{rbt}. In the maximally flippable 
ground manifold (Fig.~1), the proliferation of local loops allows spinons 
to move easily from one line to another \cite{rnbnt}, and therefore their 
motion is fully 3D. This ready availability of quantum fluctuations, 
exchanging spinon and dimer positions when $H_t$ (\ref{eht}) is applied to 
individual tetrahedra [Fig.~5(a)], means that the spinons possess quantum 
dynamics and propagate at $T = 0$. 

\begin{figure}[t]
\includegraphics[width=3.9cm]{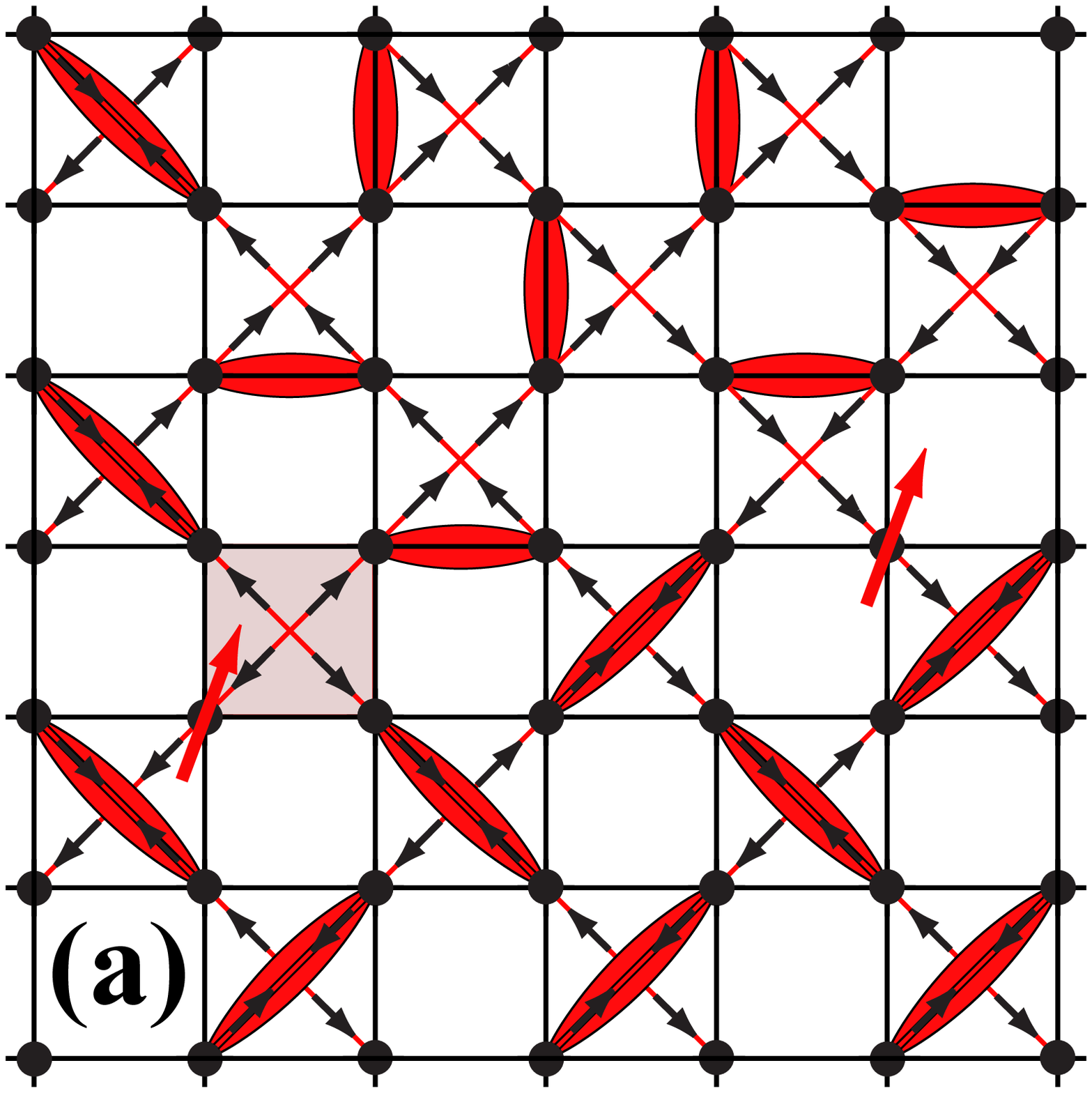}\hspace{0.2cm}\includegraphics[width=3.9cm]{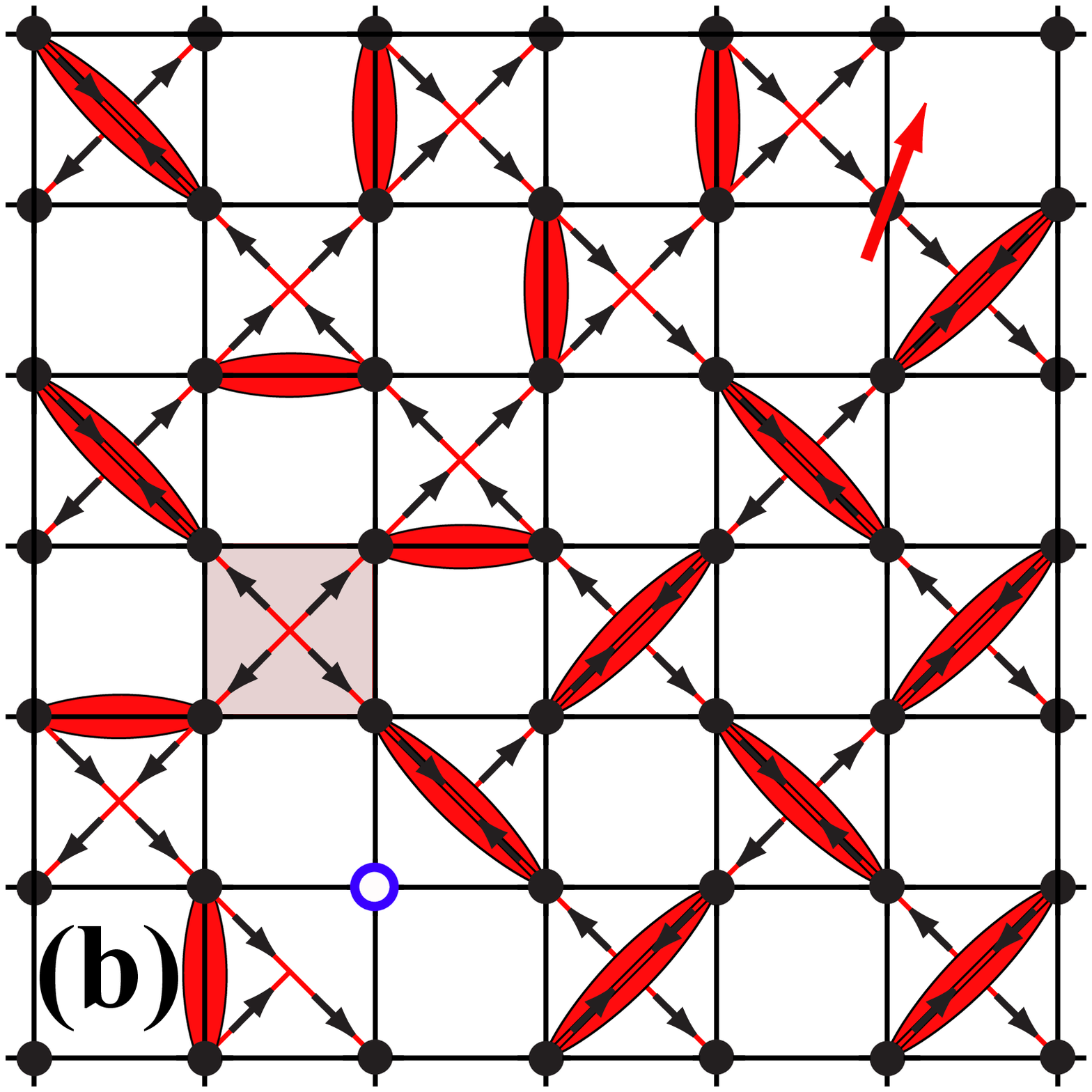} 
\caption{(color online) Representation of (a) a dissociated spinon pair 
and (b) a spin-charge-separated spinon and hole.}
\label{fig5}
\end{figure}

Charge degrees of freedom arise from a small concentration of dopants in the 
otherwise half-filled band. The energy penalty (a DT) is paid on introducing 
the hole, and the free spinon motion causes automatic spin-charge separation 
[Fig.~5(b)], leaving ``holons'' in the spin dimer background. Holon propagation 
occurs due to the kinetic term $- t \sum_{\langle ij \rangle, \sigma} c_{i\sigma}^{\dag} 
c_{j\sigma}$ in $H_{\rm {Hubb}}$ (\ref{eeh}). For such a lattice model in 3D, the 
statistics of spinons and holons can be computed from their hopping algebra 
\cite{rlw}, but this analysis requires a detailed treatment of projection 
operators describing allowed states of the spin background and lies beyond 
the scope of the current article.

It is easy to show \cite{sm} that holons experience a weak attraction to 
DTs. However, a direct binding would deny the holons the kinetic energy 
gain of propagating, albeit with a band width highly renormalized by the 
spin background. Dynamical holons experiencing a mutually attractive 
interaction by lingering close to DTs would, by these simple considerations, 
have a weak tendency towards superconductivity. This superconducting state 
is driven not by the existence of valence-bond states \cite{Anderson-1987}, 
but by the special frustration near the Klein point.

We conclude by reviewing the possibilities for finding this spin-liquid 
state in a real pyrochlore material. The Klein-point value $t/U = 1/\sqrt{30}$ 
is well within the parameter range of typical correlated insulators. 
Unfortunately, despite the wealth of pyrochlore and spinel materials 
available, very few structurally regular $S = 1/2$ systems are known. A 
fundamental property of the model (\ref{eht}) is that SU(2) spin symmetry 
is preserved, a requirement best satisfied by magnetic ions in the $3d$ series; 
to date the only candidates are the rare-earth vanadates M$_2$V$_2$O$_7$, which 
possess an additional $t_{2g}$ orbital degeneracy and are ferromagnetic. For 
pyrochlores of $4d$ ions, the SU(2) character is removed by a significant 
spin-orbit coupling. Among $5d$ ions, pyrochlore iridates have received much 
recent attention \cite{rwckb} and possess an effective $J = 1/2$ degree of 
freedom, but it is not possible to obtain effective SU(2)-symmetric 
interactions in this geometry; these materials may also be too weakly 
interacting (borderline metallic) to approach the magnetic limit. However, 
we stress again one of our key results, that energy scales for splitting of 
the degenerate manifold (Fig.~4) are very low, making spin-liquid behavior 
appear at any experimentally achievable temperatures even in systems with 
non-trivial perturbations from the Klein point. We suggest that 
pressure-dependent investigation of V$^{4+}$ and Cu$^{2+}$ materials may be 
the most promising avenue to find evidence for the spin-liquid state of the 
insulating spin-1/2 pyrochlore. 

In summary, we have demonstrated rigorously that the half-filled one-band 
Hubbard model in the pyrochlore geometry hosts a 3D quantum spin liquid. 
This spin liquid emerges, over an extended parameter regime at zero 
temperature, from a highly degenerate manifold of valence-bond states. It 
possesses massive, deconfined spinon excitations and shows spin-charge 
separation on doping. It is a quantum mechanical state essentially different 
from those studied previously, including in the 2D pyrochlore \cite{rnbnt}. 
This is one of the very few systems where unbroken degeneracies and exact 
deconfinement emerge in a realistic model with only short-range interactions. 
Finite-temperature evidence for such spin-liquid physics may be detectable 
in real pyrochlore materials.

We are indebted to C. Batista for his invaluable contributions. We thank 
R. Flint, Z.-C. Gu, Z. Hiroi, G. Ortiz, C. R\"uegg, S. Sachdev, A. Seidel, 
and T. Senthil for helpful discussions, and the Kavli Institute of Theoretical 
Physics for its hospitality. This work was supported by the NSF of China under 
Grant No.~11174365, by the National Basic Research Program of China under Grant 
No.~2012CB921704, and by the NSF under Grants CMMT 1106293 and PHY11-25915.

\section{Supplemental Material}

\subsection{S1. Klein Point}

The reason for the unique properties of the model for parameter values $t/U$  
ensuring $J_2 = - J_1$ is that the Hamiltonian $H_t$ becomes \cite{rnbnt}  
\begin{equation}
H_K = 6 J_2 \mbox{$\sum_l$} P_{S_{l,\rm tot} = 2} = {\textstyle \frac{1}{4}} 
J_2 \mbox{$\sum_l$} \vec{S}_{l,{\rm tot}}^{2} (\vec{S}_{l,{\rm tot}}^{2} - 2),
\label{esp2}
\end{equation}
with $\vec{S}_{l,{\rm tot}}^{2}  = S_{l,{\rm tot}}(S_{l,{\rm tot}} + 1)$. This is a 
sum of projectors onto the $S = 2$ sector for each tetrahedron, meaning 
that all singlet and triplet states of the four spins have energy zero and 
the only contributions are from those tetrahedra in the quintet state. Thus 
any state of the whole system with one dimer singlet, formed between any two 
of the spins, in every tetrahedron is an exact ground state of the model 
\cite{rnbnt}. This exact ground state has energy zero, as do any linear 
combinations of these states. 

This is a Klein point \cite{Klein82}. The key to its exotic properties is 
that the ground manifold has an extensive degeneracy. The proof of extensivity 
lies in the exact mappings between dimer coverings, six-vertex models, and the 
ice rules, and is known exactly in 2D \cite{baxter} and by virtue of Pauling's 
exponential lower bound in 3D \cite{rp}. Although the dimer coverings are 
not orthogonal, detailed analyses \cite{Chayes89,rs,rws2} provide strong and 
growing evidence that the number of linearly independent states in such a 
system also scales exponentially with $N$, and may even equal the number of 
dimer coverings. The most important consequence of extensive degeneracy is 
complete dimensional reduction \cite{rnbnt}, meaning that all states in the 
Klein-point manifold are connected by local (zero-dimensional) processes of 
dimer rearrangement. The proliferation of zero-energy local processes is 
responsible for the deconfinement of spinons. 

The ice rules are equivalent to a zero-divergence condition, visible in the 
arrows of the six-vertex representation (Fig.~2), which has two immediate 
consequences. One is that all correlation functions are algebraic 
\cite{baxter}, with effective dipolar correlations whose nature can be 
made intuitive in a continuum coarse-grained representation \cite{Henley}. 
The other is that, after expressing the divergence-free condition on the 
dual lattice in terms of a vector potential field, the local processes of 
dimer rearrangement are equivalent to U(1) gauge transformations, suggesting 
the nature of the associated field theory \cite{rk}. However, because every 
state is individually an eigenstate, the Klein point is a strange phase of 
matter with no quantum mechanical fluctuations \cite{rnbnt}. The different 
states are connected only at finite temperatures, by thermal fluctuations, 
and thus it is a classical critical point. The excitations created by breaking 
a dimer are two massive spinons and a defect plaquette. The availability of 
zero-dimensional local dimer fluctuations allows the spinons and defects to 
propagate freely in three dimensions. The result is a two-component Coulomb 
gas with an entropic repulsion of deconfined spinons \cite{rnbnt}. 

We comment that a subset of the concepts associated with the Klein point 
has been raised recently in the context of ``spin-ice'' pyrochlore systems 
and the experimental observation of ``magnetic monopoles'' \cite{rcms}. The 
spin-ice pyrochlore models have semiclassical values of the magnetic moments, 
with Ising anisotropies and dipolar couplings, and their asymptotically 
deconfined monopoles are in fact a more classical version of the fully quantum 
spinon excitations of the present model, which predates it \cite{rnbnt}. 
These systems are not candidate spin liquids. 

\subsection{S2. Maximally Flippable Dimer Coverings}

It is clear in 3D that all vertices of the six-vertex model are equivalent 
[Fig.~2(b)], in that every tetrahedron has four ``in-out'' edges that may 
contribute to a flippable hexagon and two (``in-in'' and ``out-out'') that 
destroy the flippability. Because every edge of a tetrahedron forms a part 
of two hexagons, it is manifestly obvious that not all hexagons in the 3D 
pyrochlore can be flippable. The difference between 2D and 3D lies in this 
fundamental geometrical property. 

To count the maximal number of flippable hexagons for any possible dimer 
covering, we consider the four interlocking kagome (111) planes of the 
pyrochlore lattice, which are shown in different colors in Fig.~1. In a 
single kagome plane, where no hexagons share an edge but do share six-vertex 
lines, at most half of the hexagons may be made flippable. It is possible to 
form lines of flippable hexagons in two planes simultaneously, shown as blue 
and green in the lower part of Fig.~1, forming a zig-zag sheet of hexagons 
sharing one pair of edges. This maximally flippable structure occupies a 
bilayer of tetrahedra [Fig.~1], all of which exploit their maximal 
flippability (four of six edges), a situation dictating a rigid dimer 
configuration. Because hexagons of the other two kagome planes, shown as 
purple and yellow, are offset from those of the blue-green pair by one 
tetrahedral unit [Fig.~1], the rigid blue-green structure excludes two 
bilayers of flippable purple and yellow hexagons. 

This offset is the key to the maximal flippable number. By leaving a single 
layer of tetrahedra above and below the blue-green bilayer, the pattern may 
then be continued with a purple-yellow bilayer, part of which is shown at 
the top of Fig.~1. A single tetrahedral layer above this is then required 
before repeating the blue-green structure. For each kagome plane, only one 
hexagon in three is flippable, but this is true for all four planes. The 
maximal number density of flippable hexagons is therefore 1/3. The bilayer 
offset is also the key to the degeneracy of the maximally flippable manifold. 
All tetrahedra in the intervening layer participate in only two flippable 
hexagons [Fig.~1] rather than four, and thus possess a two-fold degree of 
freedom. This is the orientation of the inter-bilayer singlets, or in the 
six-vertex representation the arrow direction around the hexagons of the 
upper bilayer relative to the lower. This binary choice occurs every three 
tetrahedral layers of the pyrochlore lattice. Because there are three 
primary cube axes for the layering, and there are three possible bilayers 
in the unit cell of Fig.~1, the final degeneracy is $N_f = 3$$\times$3$\times$$
2^{L/3}$, where $N = L^3/4$ is the system volume. The entropy of the maximally 
flippable manifold scales with the linear dimension $L$ and is therefore 
massive. 

\subsection{S3. Loops and Energetic Calculations}

The quantum fluctuations of a dimer covering are local dimer rearrangement 
processes, which take one dimer configuration within the ground manifold 
into another and may be represented by closed loops [Fig.~2(d)]. The 
systematic procedure to account for the effects of quantum fluctuations 
away from the Klein point is i) to identify the leading dimer-fluctuation 
loops, ii) to calculate their energetic contributions in the presence of 
the perturbation, iii) to deduce the density of contributing loops and the 
variational wave functions of the states optimizing these contributions, and 
hence iv) to obtain the degeneracy of the new manifold of states. 

\begin{table*}[t]
\caption{One-, two-, and three-hexagon dimer loops in the pyrochlore lattice.}
\begin{center}
\begin{tabular*}{15cm}{@{\extracolsep{\fill}} |c||c|c|c|c|c|c|c|c|c|c|c|c|c|}
\hline
Loop Length & 12$\,$ & 16$\,$ & 16$\,$ & 20$\,$ & 20$\,$ & 22$\,$ & 24$\,$ 
& 24$\,$ & 26$\,$ & 26$\,$ & 28$\,$ & 30$\,$ & 32$\,$ \\
\hline
$\Delta H_{ab}$  &   0 $\,$ & 0$\,$ & $\frac{\Delta J}{128} \,$ & 0 $\,$ & 
$- \frac{\Delta J}{512} \,$ & $- \frac{\Delta J}{1024} \,$ & 0 $\,$ & 
$- \frac{\Delta J}{2048} \,$ & 0 $\,$ & $- \frac{\Delta J}{4096} \,$ & 
$- \frac{\Delta J}{4096} \,$ & $- \frac{\Delta J}{8192} \,$ & 
$- \frac{\Delta J}{16384} \,$ \\
\hline
Loop Density &   1 $\,$ & $\frac{1}{2} \,$ & $\frac{1}{2} \,$ & 1 $\,$ & 1 
$\,$ & 4 $\,$ & $\frac{1}{2} \,$ & $\frac{1}{2} \,$ & 2 $\,$ & 2 $\,$ & 1 
$\,$ & 8 $\,$ & 36 $\,$ \\
\hline
\end{tabular*}
\end{center}
\end{table*}

\noindent
{\sl Loops and Loop Overlap.}
We consider the possible loops on the pyrochlore lattice (Fig.~1). Each 
loop describes the overlap of two specific states, and this can be computed 
\cite{rnbnt} from the length, number, and direction properties of each loop 
graph $G^{ab}$ as 
\begin{equation} 
g_{ab} = \langle \psi_{a}| \psi_{b} \rangle = (-1)^{n_c + L(G^{ab})/2}  
2^{N_{\Gamma}(G^{ab})- L(G^{ab})/2}, 
\label{gab} 
\end{equation} 
where $L(G^{ab}) = \sum_{j=1,N_{\Gamma}(G^{ab})} L(\Gamma^{ab}_j)$ is the total length 
of the graph, $L(\Gamma^{ab}_j)$ is the length of the loop $\Gamma^{ab}_j$ 
within it, $N_{\Gamma}(G^{ab})$ is the number of disconnected loops in the 
graph, and $n_c$ is the number of arrows with clockwise circulation around 
the loop. 

\begin{figure}[b]
\includegraphics[width=8.0cm]{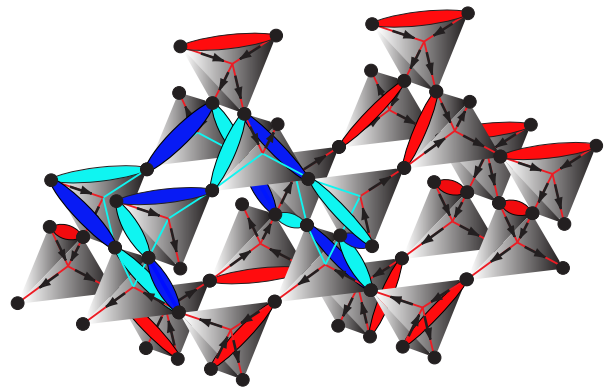}
\caption{(color online) Contributing two-hexagon, 16-bond dimer fluctuation 
process, shown as a loop of alternating light and dark blue dimers.}
\label{figs1}
\end{figure}

The available loops may be tabulated systematically by considering the 
possible connections between hexagons in the pyrochlore geometry. Beyond 
the 1-hexagon RK loop (12 bonds), two hexagons may share a side (16 bonds, 
Fig.~6), a corner (20 bonds, Fig.~7), or opposite sides of a common 
tetrahedron (22 bonds, Figs.~3 and 8). As noted in the main text, the 
majority of loops require a single (zero-energy) RK ``defect'' within the 
maximally flippable configuration. As shown in Figs.~3 and 8, the 22-bond 
loop, analogous to the 14-bond loop in Ref.~\cite{rnbnt}, is the first case 
allowing loops both with and without such an RK precursor. Loops involving 
3 hexagons may also be formed around a single defect hexagon, leading to a 
hierarchy of possibilities sharing 2 edges (20 bonds), 1 edge and 1 corner 
(24 bonds), 1 edge and 1 tetrahedron (26 bonds), 2 corners (28 bonds), 1 
corner and 1 tetrahedron (30 bonds), and 2 tetrahedra (32 bonds). The next 
families of longer loops are based on configurations involving two RK defects. 
The contributions from loops of all lengths may be calculated using the 
methods presented here. 

\begin{figure}[b]
\includegraphics[width=8.0cm]{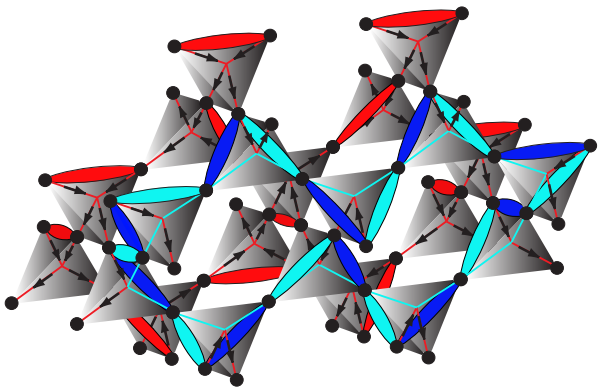} 
\caption{(color online) Two-hexagon, 20-bond dimer fluctuation process, 
shown as a loop of alternating light and dark blue dimers.}
\label{figs2}
\end{figure}

\noindent
{\sl Loop Energy Contributions.}
The calculation of matrix elements of the perturbation for each loop graph,
\begin{equation}
\Delta H_{ab} = \langle \psi_{a} |\Delta H| \psi_{b} \rangle,
\label{edhab}
\end{equation} 
exploits the fact that the Heisenberg term in $\Delta H = \sum_{ij} {\Delta J} 
\, {\bf S}_i {\bf \cdot S}_j$ is a permutation operator exchanging sites within 
a loop. Thus the energy for each pair of sites $\langle ij \rangle$ on the 
loop is a multiple of $g_{ab}$, with the prefactor taking one of only three 
possible values (2, $-1$, or 1/2) depending on the situation, and can be 
evaluated as a simple sum over all pairs \cite{rnbnt}. A key property on 
the pyrochlore lattice is that contributions from pairs of sites on the 
loop with even spacings have prefactor +2 and with odd spacings have 
prefactor $-1$; because each loop passes through three sides of an average 
tetrahedron, with two odd spacings and one even spacing along the loop, the 
contribution of a ``simple'' loop is exactly zero. Finite energetic 
contributions are therefore obtained only from loops which ``cross'' and 
therefore pass through all four sites of the same tetrahedron (Fig.~7, 
Fig.~3, Fig.~8), or which ``touch'' on two of them (Fig.~6). The matrix 
elements of $\Delta H$ for all one-, two- and three-hexagon loops are shown 
in Table II, which is a simple extension of Table I in the main text. 

\noindent
{\sl Loop Densities.}
Loops will contribute to the overall energy of the wave function according 
to their density. Two-hexagon, 16-bond loops in the maximally flippable 
configuration come in two types, both with and without the pair of vertices 
sharing a single tetrahedron visible at the top of Fig.~6, as a result of 
which only half of them have a finite energy (Table II). Two-hexagon, 20-bond 
loops have a tightly prescribed direction. For 22-bond loops, in the maximally 
flippable dimer coverings (Fig.~1), eight of the 12 hexagons coupled to any 
given flippable hexagon are also flippable, and there are two types of loop 
around every pair (Figs.~3 and 8), whence the very high density. [While 
32-bond loops have a still higher density, their energy is far lower.] 

\begin{figure}[b]
\includegraphics[width=8.0cm]{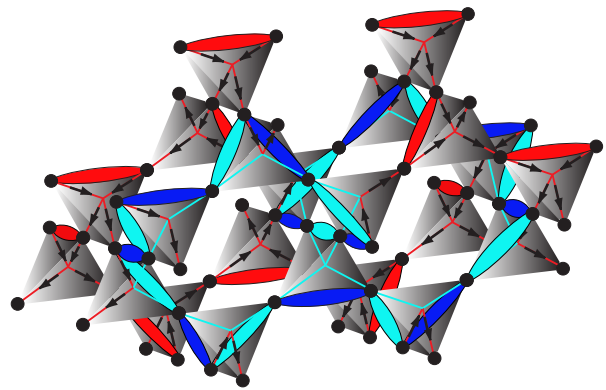} 
\caption{(color online) Two-hexagon, 22-bond dimer fluctuation process 
around the same hexagons as, but distinct from, the loop in Fig.~3.}
\label{figs3}
\end{figure}

\noindent
{\sl Ground Manifold.}
With the results of Table II it is possible to deduce the type of states 
appearing in the ground manifold. Clearly the maximally flippable states 
(Fig.~1) can only be coupled to other maximally flippable states by processes 
flipping an entire plane of spins, which from Eq.~(\ref{gab}) would give an 
exponentially small contribution to the resonance energy. The highest 
energies are gained from local loop processes, and these require the 
presence of an RK defect, a single hexagon of flipped spins that in fact 
causes four of its neighboring hexagons to become unflippable. However, any 
such state, i.e. one with $N/3 - 4$ flippable hexagons, is coupled to any 
other by all of the 16- and 20-bond loop processes and by half of the 22-bond 
loops (Fig.~3; the other half, shown in Fig.~8, connect states with $N/3$ 
flippable hexagons to states with $N/3 - 8$). Because loop energies fall 
by powers of 2 for every increment in loop length [Eq.~(\ref{gab})], states 
differing by two separate short loops contribute a much lower energy. Thus the 
ground manifold of the system is the set of maximally flippable bilayer-type 
states each with a single RK defect on any one of its $N/3$ flippable 
hegaxons, and hence $N_p = (N/3) N_f = 3N$$\times$$2^{L/3}$. While loops of all 
lengths exist, and will contribute very small corrections to the ground-state 
energy, all of the qualitative features of the optimized wave function are 
dictated by the three shortest contributing loops, because only these 
connect all the states of the ground manifold (states with $N/3 - 4$ 
flippable hexagons, which are also states differing only by the position 
of a single RK defect).

\noindent
{\sl Variational Wave Function.}
The energy gain of the most general state within this ground manifold 
of the form $|\psi \rangle = \sum_a c_a |\psi_a \rangle$ is given by 
\begin{equation}
\Delta E  = {\textstyle \frac{1}{2}} \Delta J \sum_{ l \neq k } c^{*}_{l} c_{k} 
g_{lk} \left[ {\textstyle \frac{3}{4}} L(G^{lk}) \! - \! {\textstyle \frac{1}{2}} 
|{\cal{V}}^{lk}| + W_{lk} \right], 
\label{ef}
\end{equation}
where ${\cal{V}}^{lk} = n^{lk}_{ia} + n^{lk}_{ib}$ and $W_{lk} \equiv 2 n^{lk}_{ib}
 - n^{lk}_{ia}$ are weights depending on the numbers of site pairs $\langle ij 
\rangle$ both falling on the same loop $\Gamma^{lk}$, formed by overlaying 
the singlet dimers in states $| \psi_{l} \rangle$ and $| \psi_{k} \rangle$ 
(Figs.~3, 6--8), with odd ($n^{lk}_{ia}$) or even ($n^{lk}_{ib}$) separation. The 
overall energy gain is determined from the phase structure of the coefficients 
$c_a$, which for the three shortest contributing loops can be considered as 
multiplicative factors for each of two RK-type flips $|\psi_a \rangle$ of the 
background state, linked by the loop in question. This phase pattern must 
link near-neighbor hexagons throughout the system, and cannot maximize the 
contribution of every loop simultaneously.

The dimer coverings profiting maximally from 16-, \mbox{20-,} and 22-bond 
processes are not the same, and not all virtual fluctuations can contribute 
simultaneously. The variational wave functions resulting from all of these 
loops consist of $11N/2$ superposed state pairs ($N/2$ from 16-bond loop 
processes, $N$ from 20-bond loops, and $4N$ from 22-bond loops) for each 
maximally flippable configuration. Thus there are $N_f$ such variational states 
in the ground manifold. For $\Delta J > 0$, the wave functions have all $c_a$ 
equal in absolute value but differing in sign according to the patterns shown 
in Figs.~9(a) and (b). These energetically degenerate states optimize the 
16- and 20-bond loop contributions while paying a penalty from all of the 
intra-bilayer 22-bond loops [half of the total number of these, Fig.~9(a)], 
or optimize the 16- and intra-bilayer 22-bond loop contributions while paying 
the maximum penalty for 20-bond loops [Fig.~9(b)], to gain a total resonance 
energy $\Delta E = - {\textstyle \frac{1}{256}} u^2 N \Delta J$. When $\Delta 
J < 0$, the optimal wave functions have the phase structure depicted in 
Fig.~9(c), maximizing \mbox{16-,} 20-, and intra-bilayer 22-bond loop 
contributions to gain $\Delta E = {\textstyle \frac{1}{128}} v^2 N \Delta J$. 
None of these ground-state wave functions gain energy from inter-bilayer 
fluctuations, whose contributions cancel due to the phase structure. The 
normalization coefficients $u^2$ and $v^2$ are ${\cal O} (1)$ constants 
determined from the sum of all overlaps $g_{ab}$ between the non-orthogonal 
dimer states \cite{rnbnt}.

\begin{figure}[t]
\includegraphics[width=7.5cm]{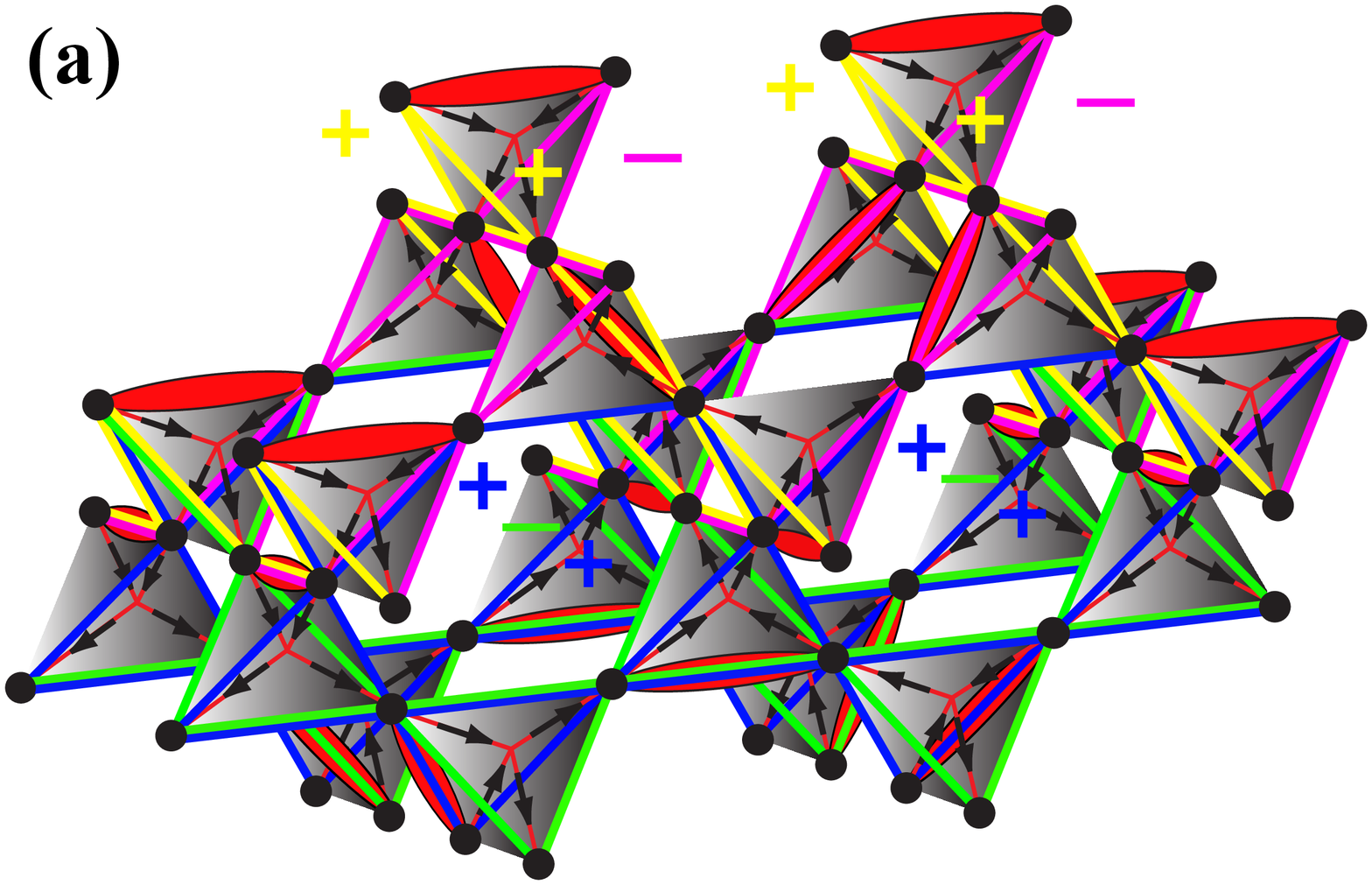}
\includegraphics[width=7.5cm]{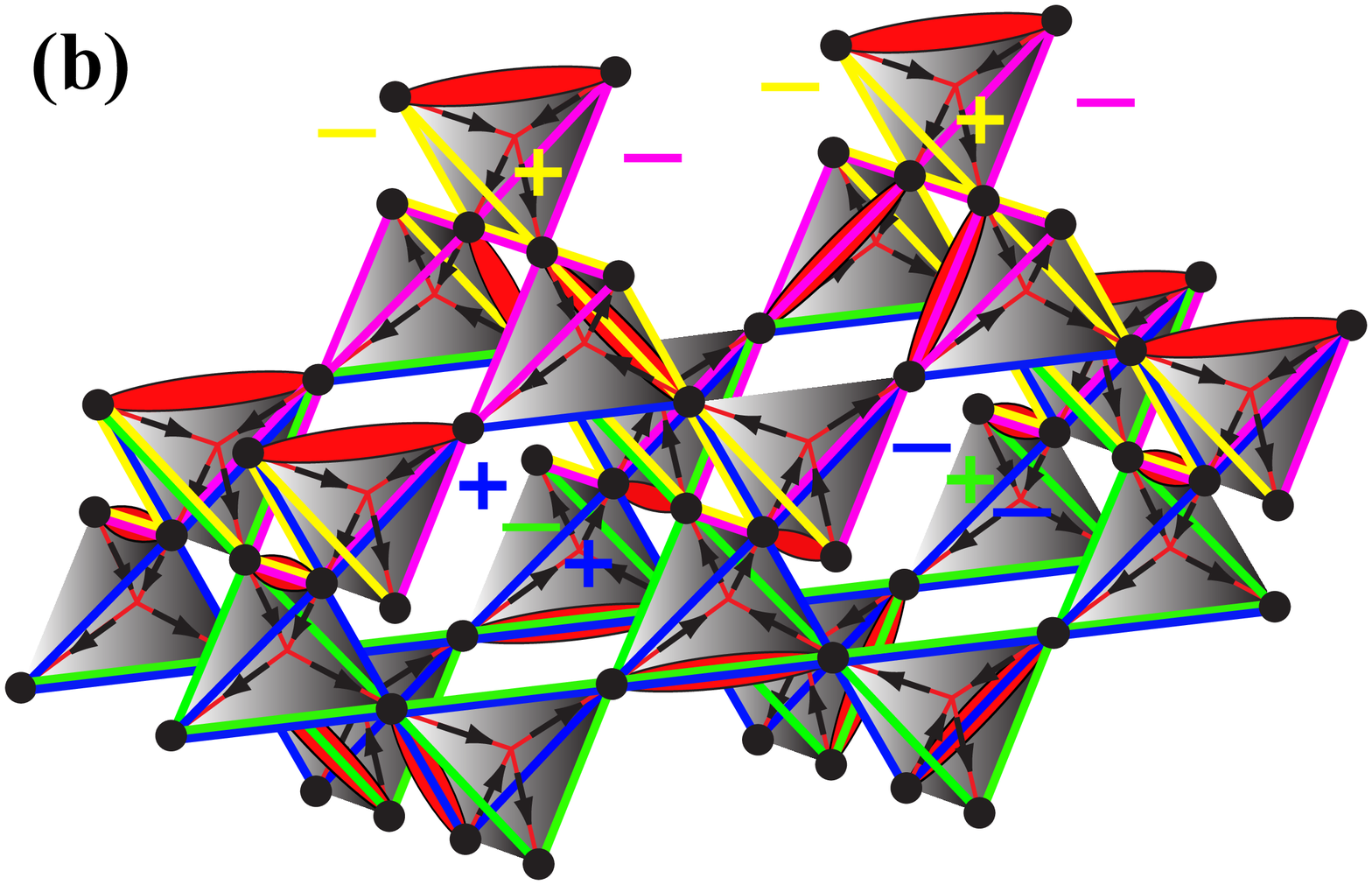}
\includegraphics[width=7.5cm]{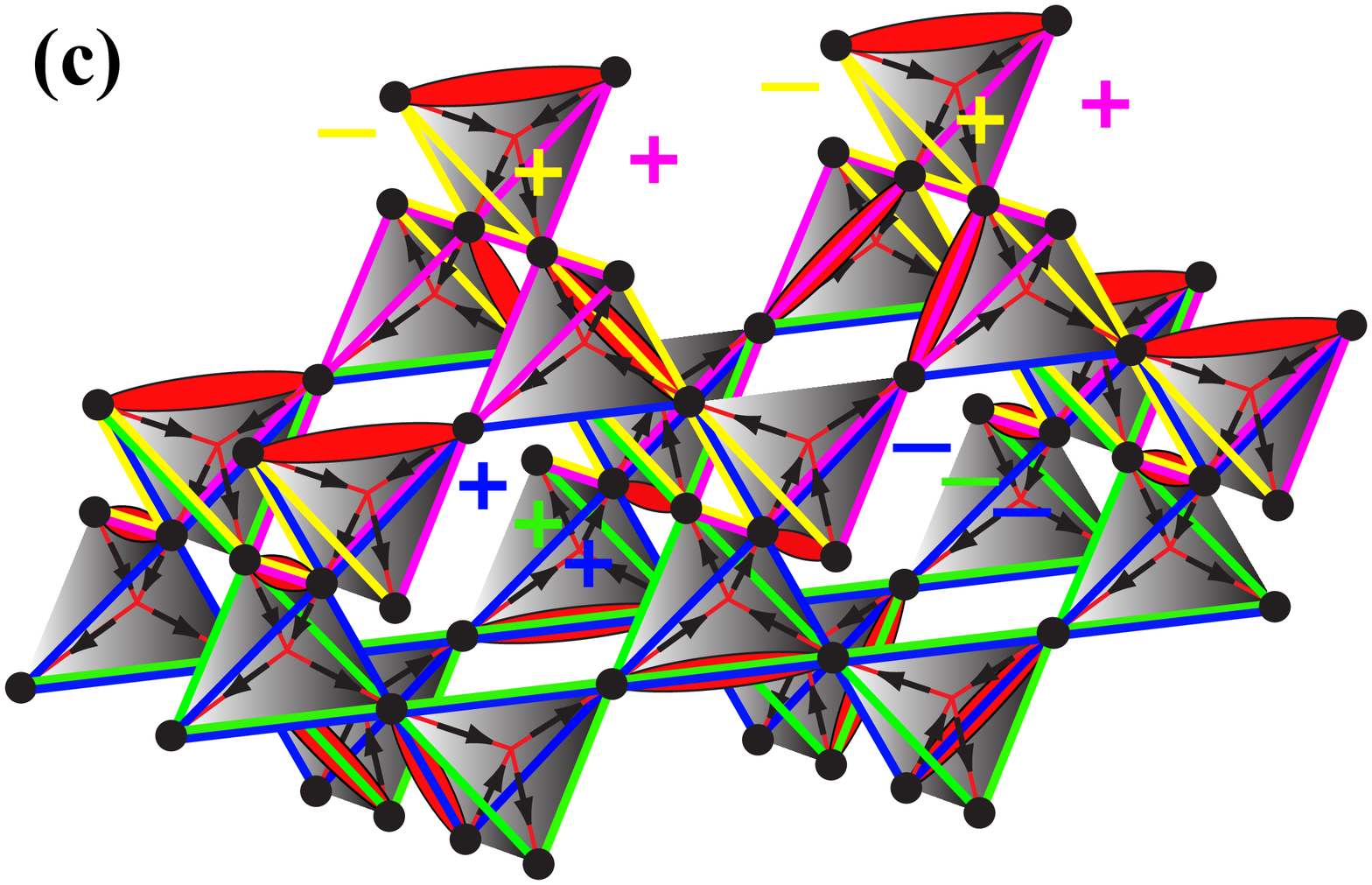} 
\caption{(color online) Phase structure of the variational wave functions 
minimizing the energy for perturbations with $\Delta J > 0$ [panels (a) and 
(b), which are degenerate] and with $\Delta J < 0$ [panel (c)]. The meaning 
of this figure is that linear superpositions of individual wave functions 
(in the overall superposition) containing single RK flips linked by 
two-hexagon loops have relative `+' or `$-$' phases in the pattern shown. }
\label{figs4}
\end{figure}

\noindent
{\sl Manifold Degeneracy.} 
Finally, it is essential to prove that the twofold bilayer degeneracies of 
the maximally flippable states cannot be lifted by any interbilayer loops, 
i.e.~that no preferred processes can lock one of the two dimer configurations 
to another in a neighboring bilayer. The 22-bond loop is the shortest one 
coupling hexagons in adjacent bilayers. For an RK defect on a hexagon in one 
bilayer, there are two possible loops to each adjacent bilayer, and both 
contribute fully for either bilayer state (direction of black arrows). No 
phase factors from the loop direction enter the calculation of the matrix 
element and no difference enters from the flippable state of either bilayer. 
The wave-function phases (Fig.~9) distinguish only between different pairs 
of hexagons, and in fact the ground states for either sign of $\Delta J$ 
lead to a total cancellation of these contributions. A different argument 
is required to include loops connecting the bilayers through two different 
tetrahedra, which go beyond any of those in Table II and thus contribute 
exponentially lower energies. These loops exist in one spatial location only 
for one of the two bilayer states, and flipping the bilayer state changes 
the location of the loop, but not its length, and hence the loop energy 
is unchanged. Further, all such loops have multiple degenerate partners 
and, because their phase structure (Fig.~9) is set by the short loops, 
their contributions always average to the same constant (usually zero for 
the ground states of Fig.~9). Combinations of these two cases cover all 
possible inter-bilayer loop types and thus the loop energies at each order 
are always the same for either flippable state of each bilayer. The 
degeneracy of the manifold of minimum-energy states therefore remains 
massive, with $N_f = 9$$\times$$2^{L/3}$ and an entropy scaling linearly with 
$L$. We remind the reader that this result emerges directly from a Klein point 
with perturbations of both intra-tetrahedron (changes to the ratio $J_2/J_1$) 
and inter-tetrahedron ($J_3$) type, covering all the leading terms in the 
minimal physical model of an insulating pyrochlore magnet. 

\subsection{S4. Spin-Liquid Nature}

The proof of exact spin-liquid nature at zero temperature requires both 
energetic and topological criteria. While several types of quantum system 
may have suitable energetic properties, there is now general agreement that 
a quantum spin liquid is a state with a specific topological order. Our 
spin model near the Klein point clearly satisfies all of the energetic 
criteria, namely the presence of a highly degenerate manifold of $N_p$ 
basis states, the energy gain from resonant processes between these states, 
and the complete spatial symmetry of the superposition of all connected 
states. While the ground manifold is determined only by specific short 
loops, resonant processes connecting all dimer coverings in this manifold 
can be represented by loops of all lengths \cite{rld}. As noted above, 
loops of length $L^2$ are required to mix the $N_f$ different states of 
the new ground manifold, and the energetic contributions from these 
processes are exponentially small. While we have cut off our explicit 
calculations at loops of length 32 (Table II), the full ground state is 
a superposition (with amplitudes $c_{a}$) of dimer states associated with 
arbitrarily long loops (and hence weak long-range entanglement), and the 
ground-state energy is a sum over all loop contributions. 

These loops hold the key to the topological criteria. The existence of 
local loop processes reflects the presence of local (formally zero-dimensional, 
or $d = 0$) gauge-type symmetries preserving the ice-rule conditions on the 
dimer arrangements. The existence of system-scale loops mixing the maximally 
flippable states corresponds to planar processes, which in turn are $d = 2$ 
gauge-type symmetries. The symmetries associated with both types of loop are 
not symmetries of the Hamiltonian, but of the ground state, and in this sense 
they are ``emergent'' at the lowest energies. The Klein-point manifold is 
split on an energy scale of order $- \Delta J / 100$ (Table II), with $N_f$ 
degenerate states at the bottom of the spectrum. These states are connected 
only by planar processes, loops with lengths of order $L^2$ in a finite 
system, which [following Eqs.~(\ref{gab}) and (\ref{ef})] have energies of 
order $- \Delta J/2^{L^2}$, and this type of exponentially small spectral gap 
is associated with topological order \cite{rwn,rno}; similar exponentially 
small differences arise also in measurements for other quasi-local operators. 
In the topologically ordered system, local loops are processes within the same 
topological sector while system-scale loops are topological operations that 
change this sector. In a finite system there are $N_f = 9$$\times$$2^{L/3}$ 
such sectors, which can be prescribed by the states of an Ising chain with 
$L/3$ sites ($L/3$ Z$_2$ variables, i.e.~an emergent ${\rm Z}_{2} \otimes 
{\rm Z}_{2} \otimes ... \otimes {\rm Z}_{2}$ symmetry). The parities of the 
bilayers in each sector and the symmetries of the processes connecting them 
determine the topological order and spin-liquid behavior, as in other known 
spin liquids including the Toric code model \cite{rak} and quantum dimer 
models \cite{rrk}. 

Indeed the situation is closely analogous to the discussion of dimer-liquid 
states in quantum dimer models, where the ground-state wave function is a 
superposition of dimer coverings with equal amplitude but differing phases, 
often referred to as the short-range resonating valence-bond (RVB) state. 
Quantum dimer models in 2D (3D) possess $d = 1$ ($d = 2$) gauge-type 
symmetries associated with the even or odd number of dimers cut by a line 
(plane) circumscribing the system, which specifies a parity, or Z$_2$ variable. 
The topological sector cannot be changed by local processes, but can be altered 
by nontrivial, system-length loop processes, and the 2 (3) independent cycles 
determine a topological degeneracy of $2^2$ ($2^3$); for comparison, in the 
present model there are $L/3$ cycles. Quantum dimer models exhibit quantum 
liquid behavior with a finite gap, Z$_2$ topological order, and fractionalized 
excitations \cite{rmr}. An extensive body of literature exists linking these 
topological concepts to the properties of RVB states in physical systems 
based on real $S = 1/2$ quantum spins, and we cite only some recent  
contributions \cite{rml,rpspgc,rspcpg,rws,Fradkin-book}. The pyrochlore 
spin model away from the Klein point provides an important additional 
dimension to these studies by presenting a demonstrably exact situation 
where a real $S = 1/2$ system possesses all of the energetic and topological 
properties of a zero-temperature quantum spin liquid.

From a practical standpoint, the issue of most relevance is the behavior 
of the system at finite temperatures, $0 < T \ll \Delta J$. An experimental 
definition of spin-liquid nature is that no local measurement can distinguish 
between ground states, in any topological sector, and thus no local probes 
may discern any type of order. Spin-liquid character can be quantified using 
the density matrix and the finite-temperature expectation values of local 
observables. Any expectation value is a Gibbs average over exponentially many 
states with infinitesimal energy splittings, whose density matrix has the form 
of an equal-amplitude sum of all admissible states in a given topological 
sector. This is precisely the canonical form of known spin-liquid wave 
functions (we cite again the examples of the Toric code and quantum dimer 
models), also at zero temperature. At finite temperatures, the expectation 
value of any quantity not invariant under all local symmetries must vanish 
by Elitzur's theorem. In this case, in contrast to the $T = 0$ case, it is 
the local gauge-type symmetries corresponding to the presence of local loop 
processes that determine the topological quantum order \cite{rno}. Further, 
when $\Delta J \ll T \ll E_{g}$, where $E_{g}$ denotes the spectral gap 
between states in the Klein-point manifold and all other states, then the 
local symmetries determined by the ice rules emerge. This is the situation 
above the dashed lines in Fig.~4. Here again, Elitzur's theorem dictates that 
only observables invariant under this increased number of local symmetries may 
attain a non-trivial expectation value.

\subsection{S5. Spinon and Holon Energies}

To explore the energetics of spinons and holons, we neglect the weak 
next-neighbor interaction ($J_3$) and begin by considering a defect 
tetrahedron (DT). Because an energy penalty is incurred only by the 
five quintet states (out of the 16 possible states) of a tetrahedron 
with four uncorrelated spins, the energy cost of a DT is $E_{\rm dt}
 = 5/16.6 J_2 = 15 J_2/8$. Spinons are uncorrelated $S = 1/2$ objects 
[Fig.~5(a)], which incur no penalty if they occupy a tetrahedron 
containing a dimer and do not change the energy of a DT if they remain 
on it. Hence the mass of one spinon is $m_s = 15 J_2/16$ and their 
motion once created is entirely free. As stated in the main text, the 
spinons are deconfined in the spin-liquid state. 

The introduction of a single hole creates a DT with a free spin, which 
may propagate freely with no energy change. A hole on a DT forms a 
tetrahedron with total spin $S_{l,{\rm tot}} = 1/2$ or 3/2 (each with 
probability 1/2), plus a neighboring tetrahedron with one dimer and 
$S_{l,{\rm tot}} = 1/2$, whence from Eq.~(\ref{esp2}) $E_{\rm bh} = [3 E 
(S_{l,{\rm tot}} = 1/2) + E (S_{l,{\rm tot}} = 3/2)]/2 = 15 J_2/32$. A hole in 
a singlet background separated from its DT creates two $S_{l,{\rm tot}} = 1/2$ 
tetrahedra and therefore has an energy $E_{\rm fh} = - 15 J_2/32$; because 
$E_{\rm dt} + E_{\rm fh} = 45 J_2/32$, there is a holon-DT binding energy 
of $15 J_2/16$. Similarly, two ``free'' holons separated from their DT 
have a total energy $E_{\rm 2fh} = 15 J_2/16$, one free and one bound holon 
have total energy $E_{\rm fh} + E_{\rm bh} = 0$, and two holons on the same 
DT have total energy $E_{\rm 2bh} = 0$. Thus the ``potential'' part of the 
effective dimer-spinon-holon Hamiltonian contains an attractive interholon 
interaction. In the event that this interaction leads to superconductivity 
of dynamical holons, if these have bosonic statistics they would condense 
individually, while fermionic holons would pair and then condense. 

\end{document}